\documentclass[superscriptaddress,showpacs,twocolumn,prl,floatfix]{revtex4-1}
\usepackage{amsfonts}
\usepackage{amsmath}
\usepackage{amssymb}
\usepackage{graphicx}
\usepackage{epstopdf}

\begin{document}

\title{Spin-orbit interaction induced singularity of the charge density relaxation propagator}

\author{S. M. Badalyan}
\email{Samvel.Badalyan@uantwerp.be}
\affiliation{Department of Physics, University of Antwerp, Groenenborgerlaan 171, B-2020 Antwerpen, Belgium}
\author{A. Matos-Abiague}
\affiliation{Department of Physics, University of Regensburg, 93040 Regensburg, Germany}
\author{J. Fabian}
\affiliation{Department of Physics, University of Regensburg, 93040 Regensburg, Germany}
\author{G. Vignale}
\affiliation{Department of Physics and Astronomy, University of Missouri-Columbia, Missouri 65211, USA}
\author{F. M. Peeters}
\affiliation{Department of Physics, University of Antwerp, Groenenborgerlaan 171, B-2020 Antwerpen, Belgium}

\date{\today}

\begin{abstract}
The charge density relaxation propagator of a two dimensional electron system, which is the slope of the imaginary part of the polarization function, exhibits singularities for bosonic momenta having the order of the spin-orbit momentum and depending on the momentum orientation. We have provided an intuitive understanding for this non-analytic behavior in terms of the inter chirality subband electronic transitions, induced by the combined action of Bychkov-Rashba (BR) and Dresselhaus (D) spin-orbit coupling. It is shown that the regular behavior of the relaxation propagator is recovered in the presence of only one BR or D spin-orbit field or for spin-orbit interaction with equal BR and D coupling strengths. This creates a new possibility to influence carrier relaxation properties by means of an applied electric field. 
\end{abstract}

\maketitle

\paragraph{Introduction} Linear response theory is one of the fundamental concepts of physics and serves as a powerful tool for studying carrier transport, relaxation and optical properties~\cite{Kubo1957,Pines1966,gv2005}. Spin-orbit interaction (SOI) modifies dramatically the carrier response~\cite{awsm2009,jf2007} allowing for a generation of spin currents when unpolarized carriers flow in single~\cite{she,murakami,kato,wunderlich2010} or bilayer~\cite{shd,sd2011} electronic systems. 
Chiral spin plasmon modes are formed due to a combined action of spin-orbit and electron-electron interaction  \cite{Ashrafi2012,Raghu2010}.  The interplay of the dominant Bychkov-Rashba (BR)~\cite{br} and Dresselhaus (D)~\cite{d} spin-orbit fields produces such fascinating effects as long-lived spatially periodic helical structures~\cite{schliemann2003,bernevig2006,koralek2009,bf2010} and magnetic spin resonances~\cite{rashbasheka,frolov2009,bf2010}, highly anisotropic propagation of electrons~\cite{flatte} and plasmons~\cite{bavf2009} with a possibility of their directional filtering.

Recently much attention has been drawn to the study of the singular response of the electron liquid with a thorough treatment of SOI effects \cite{chen,pletyukhov,simon,chesi,bfo,zak,Meng2013}. Particularly, the nonanalytical behavior of the static charge density polarizability has been exploited to predict the enhancement of RKKY interaction~\cite{simon,chesi,Meng2013} and the SOI induced beating of Friedel oscillations~\cite{bfo} leading to a more reliable quantum control of spins in potential spintronic applications. 

In contrast to the polarizability, the imaginary part of the charge density polarization function, describing dissipative properties of the near field optical response, vanishes in the static limit. However, its slope remains finite at vanishing bosonic frequencies and determines the charge density relaxation propagator, which carries additional information and in combination with the polarizability describes fully the static response of the electron system. The relaxation propagator, $K(\vec{q},\omega)$, at finite bosonic momenta, $\vec{q}$, and frequencies, $\omega$, is related to the Kubo nonlocal relaxation function, $\Psi(t, \vec{r})$, describing the system relaxation to a new equilibrium when an external force is removed at some moment \cite{Kubo1957}. The Fourier-Laplace transform of $\Psi(t, \vec{r})$ determines the relaxation propagator in terms of the charge density polarization function, $\Pi(\vec{q},\omega)$, as follows 
\begin{eqnarray}
K(\vec{q},\omega)& =&\frac{\Pi(\vec{q},0)-\Pi(\vec{q},\omega)}{i\omega}\\ \nonumber
&=&\int_{0}^{\infty} d t \int d \vec{r} \exp(-i {\bf q} {\bf r}+ i \omega t) \Psi(t, {\bf r})~.
\label{drudekernel}
\end{eqnarray}
In experiment the density relaxation propagator can be directly measured using infrared nanoscopy \cite{Fei2011, Huth2011}. Making use of super sharp tips, the current scanning probe technique reduces strongly the probing confinement region and allows characterizing the density response in the regime of large momenta and small frequencies, $v_{F}q > \omega$ ($v_{F}$ is the carrier Fermi velocity). In this regime of particular interest is the density relaxation propagator at vanishing frequencies
\begin{eqnarray}
K(\vec{q})=-\lim_{\omega\rightarrow 0} \frac{ \Im \Pi(\vec{q},\omega)} { \omega}~,
\label{drudekernel}
\end{eqnarray}
which, weighted by the impurity potential, determines the momentum relaxation rate in the Born approximation in the impurity potential~\cite{gv2005}.  

Here we calculate the charge density relaxation propagator in a two dimensional electron system in the presence of spin-orbit interaction and reveal its nonanalytical behavior. Our calculations show that $K(\vec{q})$ exhibits a singularity induced by the interplay of the BR and D spin-orbit fields. We find that the position of the singularity is given by the critical bosonic momentum, $q=q_{c}$, so that the formation of electron-hole pair excitations between the chiral Fermi contours is not possible for $q<q_{c}$ (recall that the Kohn singularity of the static polarizability occurs at much larger wave vectors $q=2k_{F}$ due to the restriction in the creation of electron-hole pairs for $q>2k_{F}$). Although the critical value of $q_{c}$ is determined by the total SOI coupling strength, the anisotropy of the energy spectrum in the presence of BR and D SOI makes it strongly dependent on the orientation, $\phi_{\mathbf{q}}$, of the bosonic wave vector $\vec{q}$, i.e. $q_{c}=q_{c}(\phi_{\mathbf{q}})$. We find that the singular behavior of $K(\vec{q})$ disappears in the limiting cases of the pure BR and pure D SOI (the overlapping form factor vanishes at $q=q_{c}$ in these isotropic cases) as well as for the SOI with equal BR and D coupling strengths (in this case, additionally, the critical momentum vanishes, $q_{c}=0$). Thus, the predicted singular behavior of the relaxation propagator, induced by the combined action of BR and D SOI, can be influenced by an external electric field and serve as a new tool for probing carrier relaxation and screening properties in experiment.

\begin{figure}[t]
\includegraphics[width=.75\linewidth]{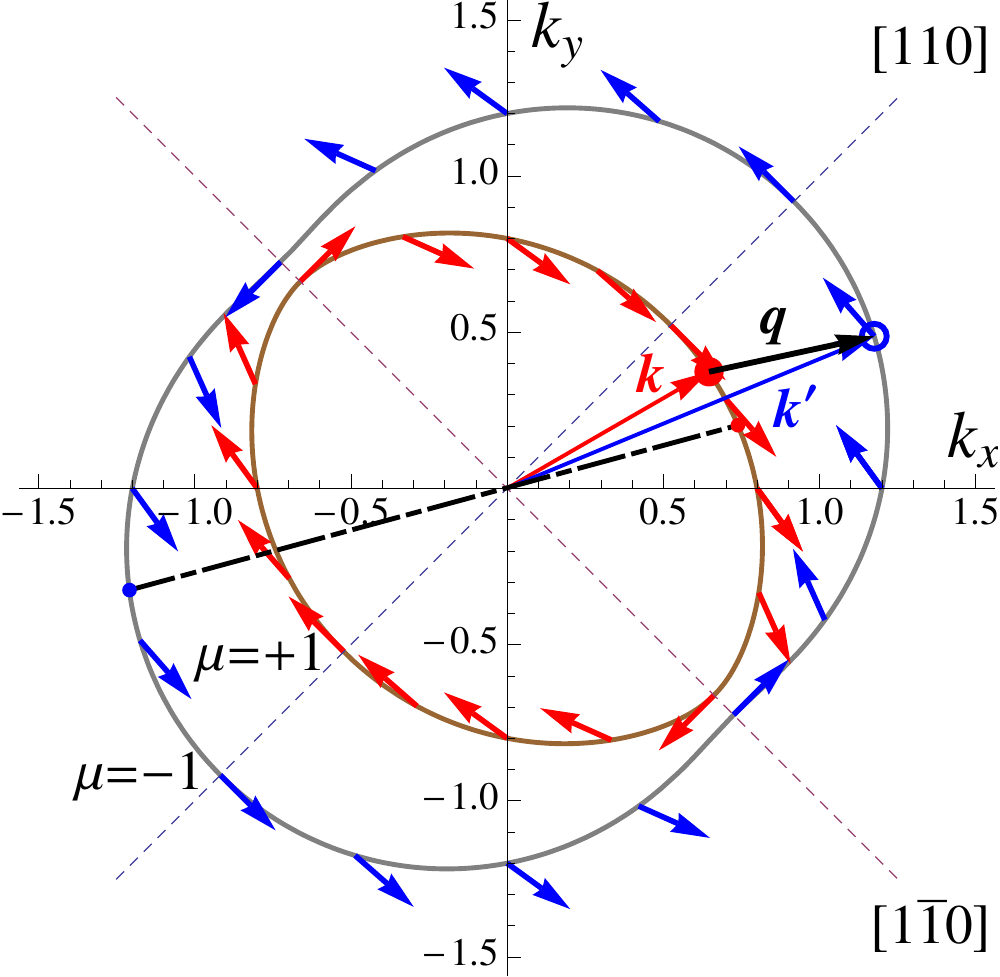}
\caption{(Color Online) The chiral Fermi contours in the presence of BR and D SOI in the $(k_{x},k_{y})$ electron momentum plane. Arrows indicate direction of the spin. A bosonic excitation of electron-hole pairs of zero energy is shown by the open and bold dots, which is mediated by carrier scattering with the initial, $\vec{k}$, and final, $\vec{k^{\prime}}=\vec{k}+\vec{q}$, momenta. Depending on the orientation of the bosonic momentum, $\vec{q}$, there exists a minimum value $q_{c}$ such that for $q< q_{c}$ it is no longer possible to form zero energy excitations on the Fermi surface, mediated by electronic transitions between states of different ($\mu=\pm1$) chirality subbands. Due to the spectrum anisotropy the form factor of overlapping spinors, in general, remains finite for scattering with $q=q_{c}$. The thick dash-dotted line is the inter chirality subband diameter,  connecting maximally distant points on the different chirality subbands along the direction of $\vec{q}$.
The ratio of BR and D SOI strengths and their absolute values are given by the parameters $\theta=\pi/5$ and $\rho =0.2k_{F}$.}
\label{fig1}
\end{figure}

\begin{figure}[t]
\includegraphics[width=.75\linewidth]{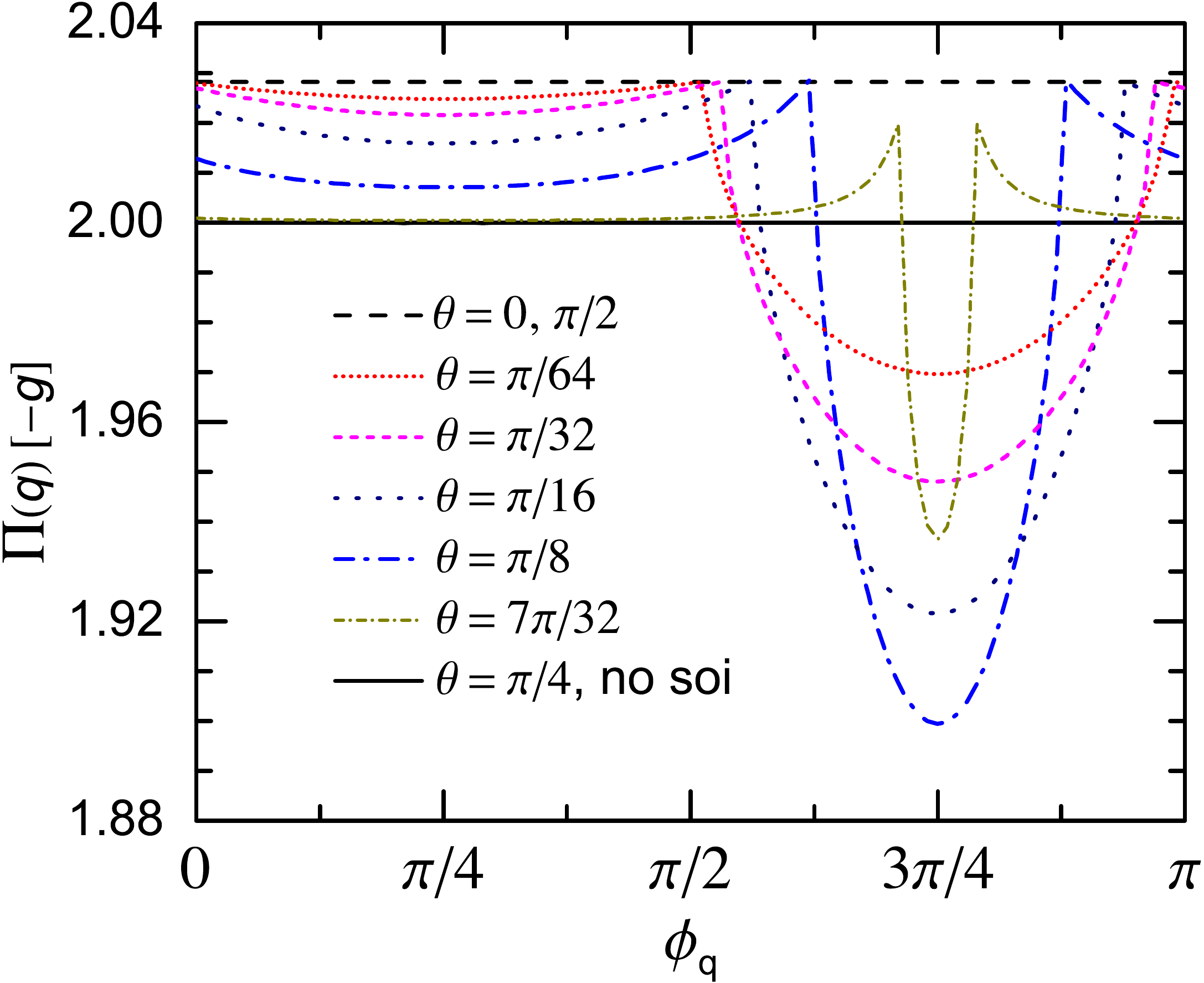} %
\caption{(Color Online) The static polarizability versus the orientation $\phi_{\mathbf{q}}$ of the bosonic momentum $\vec{q}$.  Different curves are calculated for the values of parameter $\theta$ shown in the inset and for fixed momentum magnitude, corresponding to $x=1$. Here $\rho =0.1k_{F}$.}
\label{fig2}
\end{figure}

\paragraph{Theoretical concept} The Hamiltonian of BR and D SOI in quantum wells of zinc-blende structure,
grown on a $(001)$ surface, is $H_{\text{SOI}}=\alpha \left( \hat{\sigma}_{x}k_{y}-\hat{\sigma}_{y}k_{x}\right) +\beta \left( \hat{\sigma}_{x}k_{x}-\hat{\sigma}_{y}k_{y}\right)$
where $\hat{\sigma}_{x,y}$ are the Pauli matrices, $\vec{k}$ is the in-plane electron momentum with its magnitude $k$ and polar angle $\phi _{\mathbf{k}}$. The eigenvectors of the Hamiltonian $H=H_{0}+H_{\text{SOI}}$ with $H_{0}=\vec{k}^{2}/2m^{\ast }$ ($m^{\ast }$ is the electron effective mass, $\hbar =1$) are 
$\psi _{\mu }(\vec{r})=\left(ie^{-i\varphi},\mu \right)^{T} e^{i\vec{k}\vec{r}}/\sqrt{2\mathcal{A}}$. 
They correspond to the energy branches $E_{\mu}(\vec{k})=\left[ \left( k+\mu \ \xi (\rho ,\theta ,\phi_{\mathbf{k}})\right) ^{2}-\xi (\rho ,\theta ,\phi _{\mathbf{k}})^{2}\right]/2m^{\ast }$,
which are labeled by the $\mu=\pm 1$ chirality quantum number. Here $\mathcal{A}$ is the normalization area and the phase $\varphi(\alpha ,\beta ,\phi _{\mathbf{k}})=$Arg$[\alpha e^{i\phi _{\mathbf{k}}}+i\beta e^{-i\phi_{\mathbf{k}}}]$. The angle-dependent momentum $\xi (\rho ,\theta ,\phi _{\mathbf{k}})=\rho \sqrt{1+\sin(2\theta )\sin(2\phi _{\mathbf{k}})}$ where $\rho =m^{\ast }\sqrt{\alpha ^{2}+\beta ^{2}}$ is the total SOI coupling strength. The angle parameter $\theta $ is defined as $\tan \theta =\beta /\alpha$ and describes the relative strength of the BR and D SOI. The Fermi momenta of the chirality subbands are also angle dependent, $k_{F}^{\mu }(\rho,\theta ,\phi _{\mathbf{k}})=\sqrt{2mE_{F}+\xi (\rho ,\theta ,\phi _{\mathbf{k}})^{2}}-\mu \ \xi (\rho ,\theta ,\phi _{\mathbf{k}}) $
where the Fermi energy, $E_{F}=\left( \pi n-\rho^{2}\right) /m^{\ast }$, is determined by the total carrier density $n$. Fig.~\ref{fig1} shows the anisotropic Fermi contours of the $\mu =\pm 1$ chirality subbands in the $\left( k_{x},k_{y}\right) $ plane. The inter subband scattering act, $\vec{k}\rightarrow \vec{k^{\prime}}=\vec{k}+\vec{q}$, is depicted, which mediates the formation of an electron-hole pair excitation of zero energy and with a finite wave vector $\vec{q}$.

\begin{figure*}[t]
\includegraphics[width=.32\linewidth]{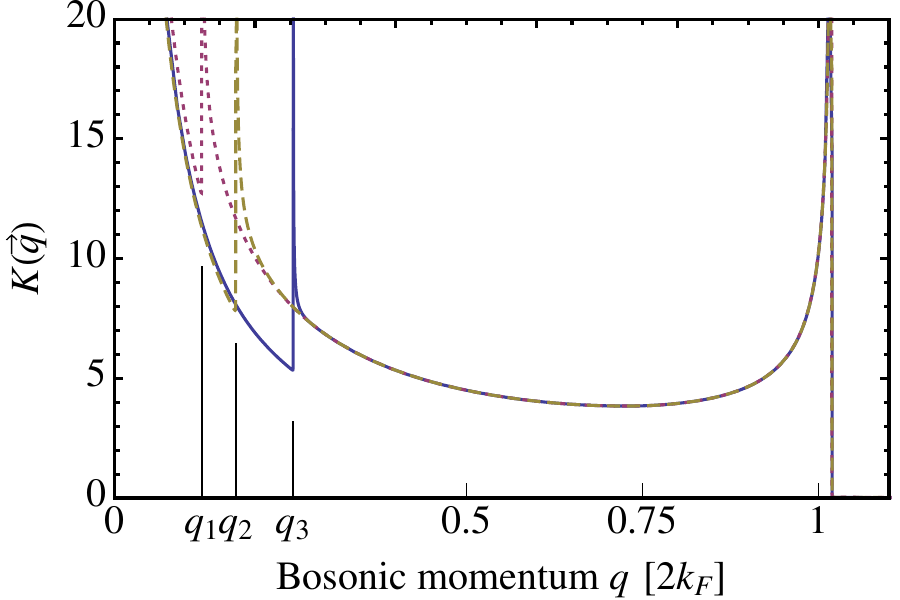} 
\includegraphics[width=.32\linewidth]{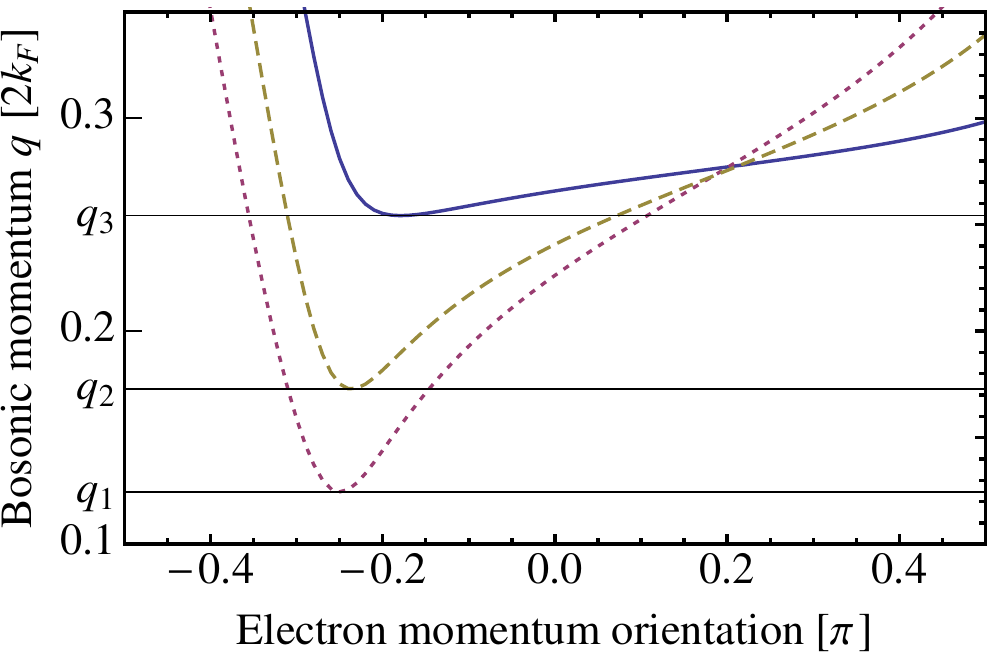}
\includegraphics[width=.32\linewidth]{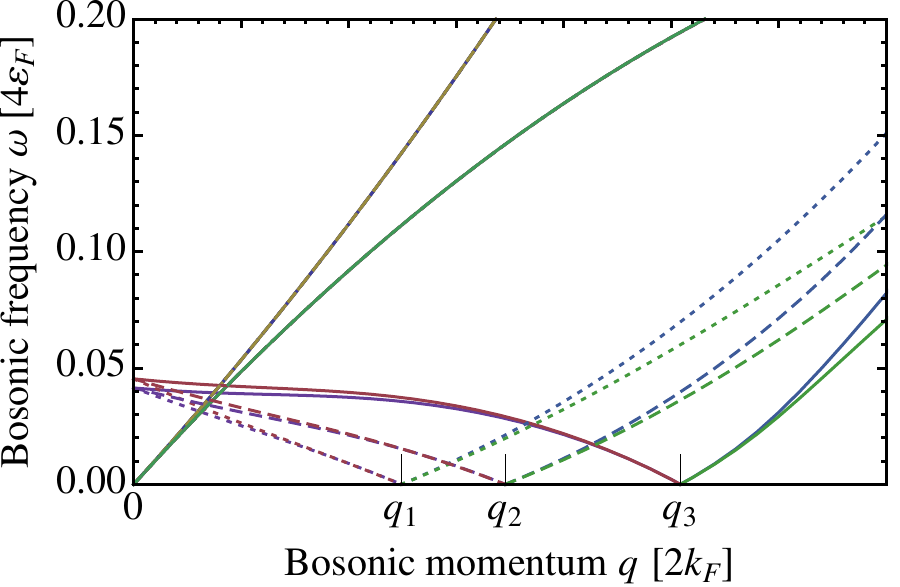}
\caption{(Color online) 
(left) The relaxation propagator $K(\vec{q})$ as a function of the bosonic momentum $q$ for three different orientations $\phi_{\mathbf{q}}=\pi /8$, $\pi/6$, and $2\pi/9$ corresponding, respectively, to the dotted, dashed, and solid curves. The positions of the singularity are shown by the large ticks, respectively, at the momentum values of $q_{c}/2k_{F}=q_{1}/2k_{F}=0.124$, $q_{2}/2k_{F}=0.173$, and $q_{3}/2k_{F}=0.254$. 
(mid) The absolute value of the bosonic momentum $q$ as a function of the electron momentum direction $\phi_{\mathbf{k}}$ for the orientations of the wave vector $\vec{q}$ corresponding to the (left) figure. 
For each curve the minimum value of $q$, shown by the gridlines at $q_{c}=q_{1}$, $q_{2}$, and $q_{3}$, gives the position of the singularity of $K(\vec{q})$. 
(right) The positions of singularities of the relaxation propagator $K(\vec{q},\omega)$ in the ($\omega,q$) plane. The solid curves starting from the origin show the singularities at the boundaries of the intra chirality subband electron-hole continuum. The format of other curves in the (right) figure and the curves in the (mid) figure corresponds to the same values of $\phi_{\mathbf{q}}$ in the (left) figure. The SOI parameters used here are the same as in Fig.~\ref{fig1}.}
\label{fig3}
\end{figure*}

In Eq.~(\ref{drudekernel}) the polarization function in the presence of the BR and D SOI of arbitrary strengths is a sum over the chirality indices, $\Pi (\vec{q},\omega )=\sum_{\mu,\nu =\pm 1}\Pi _{\mu \nu }(\vec{q},\omega )$, with
\begin{eqnarray}
\Pi_{\mu \nu }(\vec{q},\omega ) &=&\int \frac{d\vec{k}}{\left( 2\pi \right)^{2}}\frac{f(E_{\mu }(\vec{k}))-f(E_{\nu }(\vec{k}+\vec{q}))}{E_{\mu }(\vec{k})-E_{\nu }(\vec{k}+\vec{q})+\omega +i0}  \label{pf} \\
&&\times \mathcal{F}_{\mu \nu }\left( \vec{k},\vec{k}+\vec{q}\right) ~. \notag
\end{eqnarray}
where $f(E_{\mu }(\vec{k}))$ is the Fermi distribution function. The form factor $\mathcal{F}_{\mu \nu }\left( \vec{k},\vec{k}+\vec{q}\right)=\left[ 1+\mu \nu \cos \left( \Delta \varphi _{\mathbf{q}}\right) \right]/2$ determines the overlap of the spinor wave functions of scattered particles.  
Here we define $\Delta \varphi _{\mathbf{q}}=\varphi (\alpha ,\beta ,\phi _{\mathbf{k}})-\varphi (\alpha ,\beta ,\phi _{\mathbf{k}+\mathbf{q}})$. Taking analytically the integration over $k$ in (\ref{pf}),  we reduce the polarization function to an average over the electron polar angle as
\begin{eqnarray}
\Pi (\vec{q},\omega )&=&-\frac{g}{4\pi }\sum_{\mu,\lambda=\pm1}\int_{0}^{2\pi}d\phi_{\mathbf{k}} P_{\mu,\lambda}(\vec{q},\omega|\phi_{\mathbf{k}})
\label{av}
\end{eqnarray}
with $g=m^{\ast }/2\pi $ denoting the density of states at the Fermi level and
\begin{eqnarray}
&&P_{\mu,\lambda}(\vec{q},\omega|\phi_{\mathbf{k}})= \frac{v_{F}d}{a}+\frac{1}{a\left(v^{+}-v^{-}\right)}
\label{intgrd} \\
&\times&
\left[ v^{+}(e-dv^{+})\ln \frac{v^{+}}{v^{+}-v_{F}}-v^{-}(e-dv^{-})\ln \frac{v^{-}}{v^{-}-v_{F}}\right]~. \notag
\end{eqnarray}
We have introduced the dimensionless Fermi wave vector $v_{F,\mu }=\sqrt{1-r^{2}+\overline{\xi }_{\mathbf{k}}^{2}}-\mu \overline{\xi }_{\mathbf{k}}$ and the functions $v_{\mu ,\lambda }^{\pm }=\left( -b_{\mu ,\lambda }\pm 
\sqrt{{b_{\mu ,\lambda }^{2}}-4a_{\mu }c_{\lambda }}\right) /2a_{\mu }$ together with the coefficients
\begin{subequations}
\begin{eqnarray}
a_{\mu } &\equiv & x\cos \left( \phi _{\mathbf{k}}-\phi _{\mathbf{q}}\right)\left[ x\cos \left( \phi _{\mathbf{k}}-\phi_{\mathbf{q}}\right) -\mu \overline{\xi }_{\mathbf{k}}\right] ~,  \label{eq10} \\
b_{\mu ,\lambda } &\equiv &-x\left[ \left( r^{2}+2(\lambda y-x^{2})\right)\cos \left( \phi _{\mathbf{k}}-\phi _{\mathbf{q}}\right) \right.
\label{eq11} \\
&+&\left. r^{2}\sin \left( 2\theta \right) \sin \left( \phi _{\mathbf{k}}+\phi _{\mathbf{q}}\right) \right] +\mu (\lambda y-x^{2})\overline{\xi }_{\mathbf{k}}~,  \notag \\
c_{\lambda } &\equiv &\left( \lambda y-x^{2}\right) ^{2}-x^{2}\overline{\xi }_{\mathbf{q}}{}^{2}~,  \label{eq12} \\
d_{\mu } &\equiv &x\cos \left( \phi _{\mathbf{k}}-\phi _{\mathbf{q}}\right)-\mu \overline{\xi }_{\mathbf{k}}~,  
\label{eq13} \\
e_{\mu ,\lambda } &\equiv &\lambda y-x^{2}  \label{eq14} \\
&+&\frac{\mu r^{2}x}{\overline{\xi }_{\mathbf{k}}}\left[ \cos \left( \phi _{\mathbf{k}}-\phi _{\mathbf{q}}\right) +\sin \left( \phi _{\mathbf{k}}+\phi _{\mathbf{q}}\right) \sin (2\theta )\right] ~.  \notag
\end{eqnarray}
\end{subequations}
We use the following dimensionless quantities $x=q/2k_{F}$, $y=\omega/4\varepsilon _{F}+i0$, $v=k/k_{F}$, $r=\rho k_{F}$, and $\overline{\xi }_{\mathbf{k}}=\xi (\rho ,\theta ,\phi _{\mathbf{k}})/k_{F}$ with $\varepsilon_{F}=k_{F}^{2}/2m^{\ast }$ and $k_{F}=\sqrt{2m^{\ast }E_{F}+\rho ^{2}}$~. For brevity, on the rhs of Eq.~(\ref{intgrd}) we have omitted the indices $\mu$ and $\lambda$.

\paragraph{Results and discussion} 

In Fig.~\ref{fig2} we demonstrate the strongly anisotropic behavior of the static polarization function $\Pi (\vec{q})$,  plotting its dependence on the bosonic momentum orientation $\phi _{\mathbf{q}}$ for several values of the relative strength of the BR and D SOI given by the parameter $\theta$ and for fixed momentum magnitude corresponding to $q=2k_{F}$. The two singular points of the polarization function separates the $\phi _{\mathbf{q}}$ range into regions where $\Pi (\vec{q})$ behaves qualitatively different. Depending on whether or not electronic transitions on the chiral Fermi contours mediate bosonic electron-hole excitations with zero energy and finite momenta, the character of variation of $\Pi (\vec{q})$ is, respectively, smooth or with a larger amplitude. Notice that the combined effect of BR and D SOI with equal strengths ($\theta =\pi /4$ or $3\pi /4$) cancels each other and we have $\Pi (\vec{q})=2$ for all orientations of the bosonic momentum.

\begin{figure}[t]
\includegraphics[width=.85\linewidth]{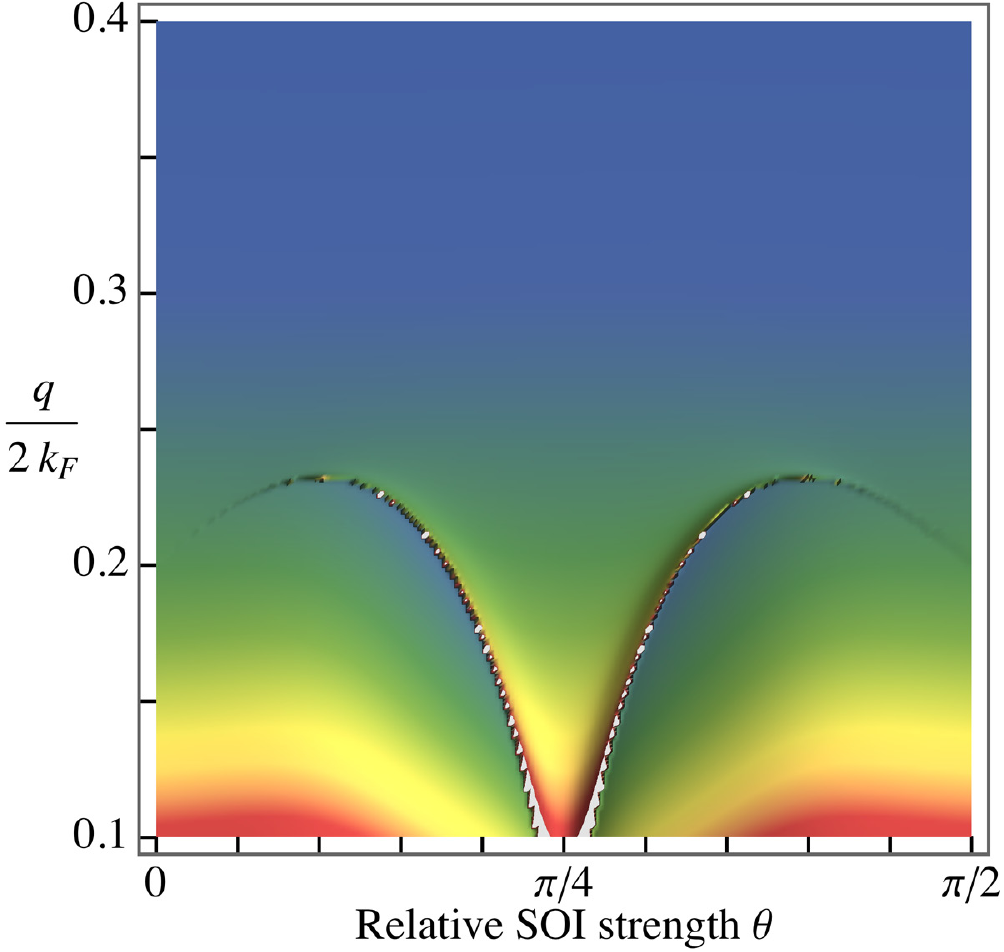} 
\caption{(Color online) Density plot of the charge density relaxation propagator $K(\vec{q})$ as functions of the parameter $\theta$ and of the bosonic momentum $q/2k_{F}$. The total SOI strength $\rho=0.2$ and the orientation of the momentum $\phi_{\mathbf{q}}=\pi/6$. The color output plot range is from 5.3 (blue) to 20.5 (red).} 
\label{fig4}
\end{figure}

In Figs.~\ref{fig3} and \ref{fig4} we study the singular behavior of the charge density relaxation propagator. In Fig.~\ref{fig3}(left) we plot the relaxation propagator as a function of the bosonic momentum $q$ for three different orientations, $\phi_{\mathbf{q}}=\pi /8$, $\pi/6$, and $2\pi/9$. In addition to the conventional singularities at vanishing $q$ and at large momenta near $2k_{F}$, the relaxation propagator exhibits a small-$q$ singularity. The position of this new singularity, $q_{c}$, is determined by the total SOI coupling strength $\rho$. As seen, $q_{c}$ depends on the orientation of the wave vector $\vec{q}$. To identify the origin of the relaxation propagator singularity in terms of the single particle electronic transitions, we plot in Fig.~\ref{fig3}(mid) the dependence of the absolute value $q$ of the bosonic momentum as a function of the polar angle $\phi_{\mathbf{k}}$ of the electron momentum $\vec{k}$ for the respective orientations of $\vec{q}$ from Fig.~\ref{fig3}(left). At the same time it is assumed that the wave vectors $\vec{k}$ and $\vec{k^{\prime}}$ remain, respectively, on the Fermi contours with the chirality $\mu=+1$ and $\mu=-1$. It is seen in Fig.~\ref{fig3}(mid) that the three curves, corresponding to the three different orientations of $\vec{q}$, exhibit clear minima at some values of $\phi_{\mathbf{k}}$ and the obtained minimal values of $q$, labeled as $q_{c}=q_{1}$, $q_{2}$, and $q_{3}$, determine the positions of the respective singularities of $K(\vec{q})$ in Fig.~\ref{fig3}(left). The existence of the singularity reflects the fact that the formation of {\it inter chirality} electron-hole bosonic excitations of zero energy is no longer possible for $q<q_{c}$. In contrast, the large-$q$ singularity of $K(\vec{q})$ in Fig.~\ref{fig3}(left) is determined by the length, $q_{max}$, of the diameter of the chiral Fermi contours shown in Fig.~\ref{fig1} by the  thick dash-dotted line. This is because inter chirality subband electronic transitions on the Fermi contours are not possible for $q>q_{max}$. The dependence of  $q_{max}$ on the orientation of $\vec{q}$ for these values of parameters is so weak that the respective changes of the peak position are not visible on the scale of Fig.~\ref{fig3}(left). In Fig.~\ref{fig3}(right) we study the SOI induced singularity of the relaxation propagator at finite frequencies, where we show the critical momenta $q_{c}=q_{1}$, $q_{2}$, and $q_{3}$ in the $(\omega, q)$ plane along the boundaries of the inter chirality subband particle-hole continuum as induced by the interplay of BR and D SOI. Notice that at small values of frequencies there exists also particle-hole continuum formed due to {\it intra chirality subband} electronic transitions so that the relaxation propagator exhibits singularities also along the respective boundaries shown by the solid curves starting from the origin. 

Fig.~\ref{fig4} shows a density plot of the relaxation propagator $K(\vec{q})$ with the parameter $\theta$, describing the relative strength of the BR and D SOI, and with the absolute value of the bosonic momentum $q$ for its orientation fixed at $\phi_{\mathbf{q}}=\pi/6$. The singularity of $K(\vec{q})$ for vanishing values of $q$ is not shown. Although it is stronger than the SOI induced small-$q$ singularity, its contribution to such an important physical quantity as the momentum relaxation rate 
is suppressed by an additional factor of $q^{2}$~\cite{gv2005}, which weights the relaxation propagator. 
The singularity of the relaxation propagator disappears at finite values of the bosonic momenta $q$ for equal BR and D SOI coupling strengths ($\theta=\pi/4$). As seen in Fig.~\ref{fig4} the singularity disappears smoothly also in the limits of $\theta=0$ or $\pi/2$ corresponding, respectively, to the cases of pure BR SOI or pure D SOI. Thus, the BR and D SOI induced anisotropy of the single-particle spectrum is responsible for a finite overlap of the electron and hole spinor wave functions at $q=q_{c}$ and thereby for the appearance of the small-$q$ singularity of the relaxation propagator.

%

In conclusion, we predicted a singular behavior of the charge density relaxation propagator, which is an important density response function of a quantum gas of electrons. An intuitive understanding has been provided for this small-$q$ nonanalyticity in terms of electronic transitions between the chiral electronic subbands induced by the combined action of BR and D spin-orbit fields. The relaxation propagator recovers its regular behavior in the limiting cases of pure BR and pure D SOI and in spin-orbit fields of equal strengths. This puts forward a new mechanism to tune electrically the carrier relaxation properties by adjusting the relative SOI strength.

We acknowledge support from the Methusalem program of the Flemish government and the Flemish Science Foundation (FWO-Vl), DFG SFB Grant 689, and NSF Grant DMR-1104788 (G.V.). 


\end{document}